\documentclass[amsfonts, amsmath, amssymb, aps, pra, twocolumn, a4paper, superscriptaddress, floatfix]{revtex4-2}

\usepackage[toc,page,title,titletoc,header]{appendix}
\usepackage[T1]{fontenc}
\usepackage{enumerate}
\usepackage{graphicx}
\usepackage[latin1]{inputenc}
\usepackage{epstopdf}
\usepackage{amsthm}
\usepackage{amsmath}
\usepackage{csquotes}
\usepackage{bbm}
\usepackage{gensymb}
\usepackage{amsmath}
\usepackage{mathptmx}
\usepackage{natbib}
\usepackage[misc]{ifsym}
\usepackage{physics}
\usepackage{subfig}
\usepackage{multirow}
\usepackage{xcolor}
\usepackage{ragged2e}
\DeclareCaptionJustification{justified}{\justifying}
\usepackage[skip=2pt,font=footnotesize]{caption}
\renewcommand{\thefigure}{\arabic{figure}}
\renewcommand{\thetable}{\arabic{table}}
\usepackage{amssymb, amsbsy, amsthm}
\usepackage{times} 

\usepackage{fancyhdr}
\usepackage{lineno}

\begin{document}

\title{\LARGE \textbf{Molecular property prediction \\with photonic chip-based machine learning}}

\author{Hui Zhang}
\thanks{These two authors contributed equally}
\affiliation{Quantum Science and Engineering Centre (QSec), Nanyang Technological University, 50 Nanyang Ave, 639798 Singapore, Singapore}
\author{Jonathan Wei Zhong Lau}
\thanks{These two authors contributed equally}
\affiliation{Centre for Quantum Technologies, National University of Singapore, 3 Science Drive 2, Singapore 117543}
\author{Lingxiao Wan}
\affiliation{Quantum Science and Engineering Centre (QSec), Nanyang Technological University, Singapore}
\author{Liang Shi}
\affiliation{Chemistry and Biochemistry, University of California, Merced, California 95343, United States}
\author{Yuzhi Shi}
\affiliation{Institute of Precision Optical Engineering, School of Physics Science and Engineering, Tongji University, Shanghai 200092, China}
\author{Hong Cai}
\affiliation{Institute of Microelectronics, A*STAR (Agency for Science, Technology and Research), Singapore}
\author{Xianshu Luo}
\affiliation{Advanced Micro Foundry, 11 Science Park Rd, Singapore}
\author{Guo-Qiang Lo}
\affiliation{Advanced Micro Foundry, 11 Science Park Rd, Singapore}
\author{Chee-Kong Lee}
\email[Corresponding Author: ]{cheekonglee@tencent.com (C.K.L), cqtklc@gmail.com (L.C.K), eaqliu@ntu.edu.sg (A.Q.L)}
\affiliation{Tencent America, Palo Alto, CA 94306, United States}
\author{Leong Chuan Kwek}
\email[Corresponding Author: ]{cheekonglee@tencent.com (C.K.L), cqtklc@gmail.com (L.C.K), eaqliu@ntu.edu.sg (A.Q.L)}
\affiliation{Quantum Science and Engineering Centre (QSec), Nanyang Technological University, Singapore}
\affiliation{Centre for Quantum Technologies, National University of Singapore, Singapore}
\affiliation{National Institute of Education, Nanyang Technological University, Singapore}
\affiliation{MajuLab, CNRS-UNS-NUS-NTU International Joint Research Unit, UMI 3654, Singapore}
\author{Ai Qun Liu}
\email[Corresponding Author: ]{cheekonglee@tencent.com (C.K.L), cqtklc@gmail.com (L.C.K), eaqliu@ntu.edu.sg (A.Q.L)}
\affiliation{Quantum Science and Engineering Centre (QSec), Nanyang Technological University, Singapore}

\begin{abstract}
Machine learning methods have revolutionized the discovery process of new molecules and materials. However, the intensive training process of neural networks for molecules with ever-increasing complexity has resulted in exponential growth in computation cost, leading to long simulation time and high energy consumption. Photonic chip technology offers an alternative platform for implementing neural networks with faster data processing and lower energy usage compared to digital computers. Photonics technology is naturally capable of implementing complex-valued neural networks at no additional hardware cost. Here, we demonstrate the capability of photonic neural networks for predicting the quantum mechanical properties of molecules. To the best of our knowledge, this work is the first to harness photonic technology for machine learning applications in computational chemistry and molecular sciences, such as drug discovery and materials design. We further show that multiple properties can be learned simultaneously in a photonic chip via a multi-task regression learning algorithm, which is also the first of its kind as well, as most previous works focus on implementing a network in the classification task.
\end{abstract}

\maketitle
\section{Introduction}
Data-driven approaches, in particular machine learning methods, have in the past decade become an indispensable tool for material design and molecular discovery~\cite{Butler2018, VonLilienfeld2020}. The availability of large datasets, both experimental and computational, and the advancements in algorithms and computer capability significantly accelerate the screening and identification of molecules and materials with desired properties in vast chemical space. 
Among the machine learning methods, deep learning methods involving large and sophisticated neural networks have emerged as the leading candidates for molecular property prediction. 
However, despite their promising performance, training of such deep neural networks inadvertently leads to substantial computational costs associated with expensive hardware and high power consumption. This is fundamentally due to the fact that modern computers are built on the von Neumann architecture, and training neural networks is a very memory-intensive task, which is then subjected to the von Neumann bottleneck~\cite{zou2021breaking}.
For example, the highly successful recent work, Alphafold~\cite{jumper2021highly}, for protein folding typically requires training times in the order of weeks using hundreds of costly graphic processing units (GPUs)~\cite{deepmind}. Furthermore, while the size and computational costs of the largest machine learning models continue to double every few months~\cite{amodei_2021}, the efficiency of digital computational chips has not managed to improve at a similar pace~\cite{reuther2020survey}. Thus, training of ever more sophisticated machine learning models is expected to lead to the exponential growth of energy consumption~\cite{Henderson2020}, which comes with a huge environmental cost~\cite{strubell2019energy}.

Optical computing, with its ability to perform arbitrary linear functions between inputs and outputs~\cite{miller2013self}, offers an alternative paradigm to conventional digital computing for computation-intensive tasks, such as implementing deep neural networks~\cite{zhou2022photonic}.
Optical computing implementations of machine learning methods have been realized in neuromorphic photonics~\cite{de2019machine,tait2017neuromorphic}, all-optical neural networks~\cite{feldmann2019all,lin2018all} and photonic reservoir computing~\cite{van2017advances,vandoorne2014experimental}. 
In fact, optical implementation of neural networks has recently been demonstrated to surpass cutting-edge GPUs in terms of speed and energy consumption, and is potentially able to achieve orders of magnitude improvements over conventional neural networks for standard problem sizes~\cite{shen2017deep}.
Optical computing offers several advantages over conventional digital computers, such as low power usage~\cite{caulfield2010future,ahn2009devices,tait2021quantifying}, ultrafast optoelectronics with high noise robustness~\cite{prucnal2016recent}, inherent parallelism~\cite{larger2012photonic}, and large information storage~\cite{hill2004fast}. Schemes have shown that all these advantages are scalable to large problems~\cite{hamerly2019large}.
Furthermore, optical neural networks implementing complex-valued neural networks have been developed recently~\cite{zhang2021optical}. 

Despite the promises of the optical neural network, its potential in molecular sciences such as chemistry and biology has not been explored. Additionally, most current works on optical neural networks focus on the single-task learning (STL) method. Yet, humans typically acquire knowledge and arrive at generalization through multi-task learning (MTL). In the study of molecules, we often encounter the need to predict the values of many properties (e.g. free energy and enthalpy) as the most desirable molecules or materials typically optimize multiple properties simultaneously. This sort of multi-task learning differs fundamentally from single-task classification and requires modified algorithms. Instead of producing a category code to categorize a set of input data into a specific class, we seek to produce and optimize a continuous real-valued output~\cite{goodfellow2016deep}. It is also desired to learn multiple attributes (tasks) at the same time, in the hope that we can leverage the knowledge contained in one task to generalize and improve the performance of other tasks~\cite{caruana1997multitask,zhang2017survey}. With this motivation in mind, we hope to implement multi-task regression neural network within the optical scheme. Furthermore, molecular properties are inherently quantum mechanical and the calculations involve complex-valued arithmetics, it would therefore be appealing to attempt molecular property prediction with complex-valued neural networks, a function that can be accomplished on the same optical chip with no additional hardware cost.

\begin{figure*}
    \centering
    \includegraphics[width=0.72\textwidth]{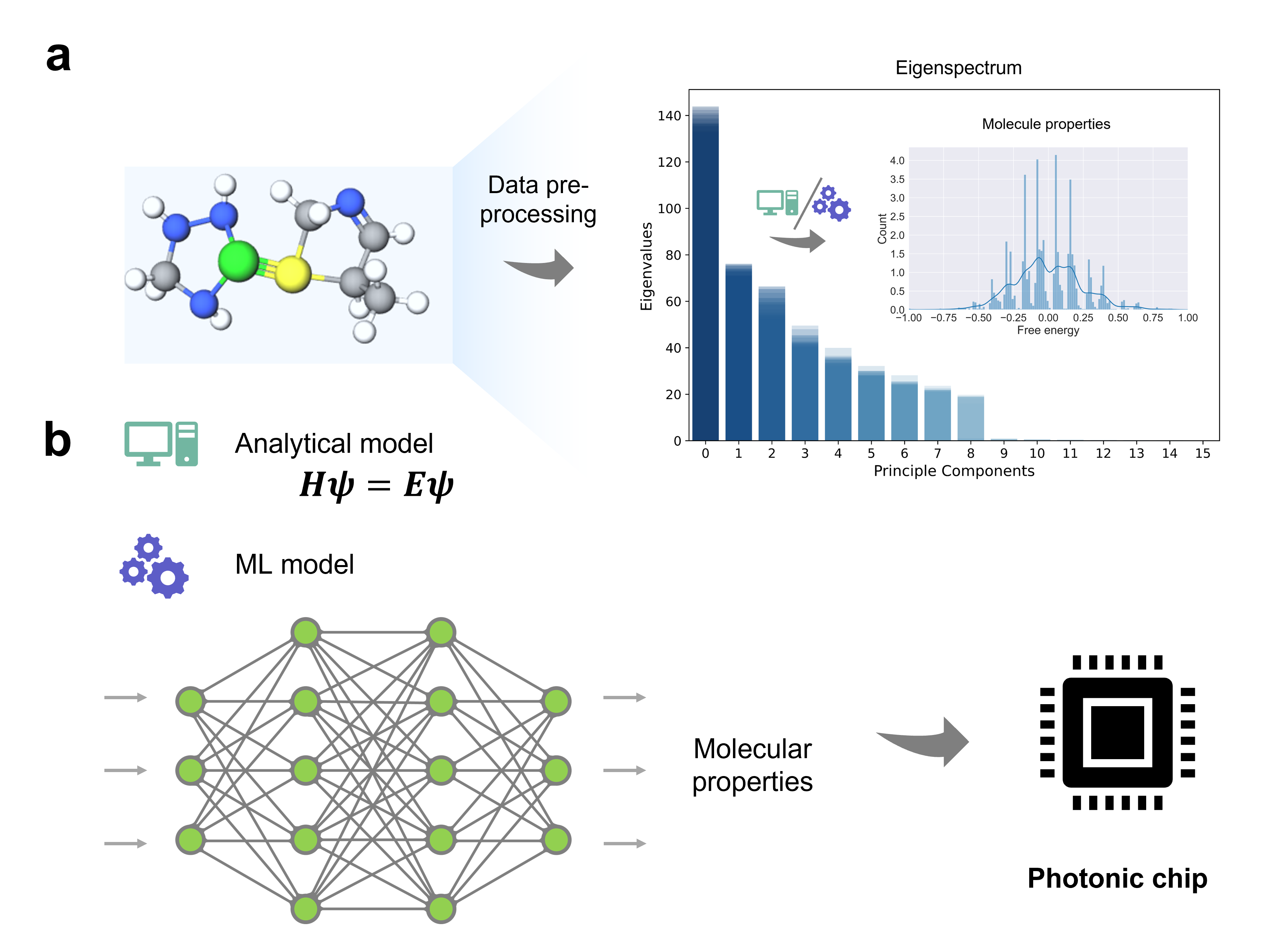}
    \caption{Task briefs and illustrations of the machine learning method for predicting molecular properties. \textbf{a}, The task of predicting molecular properties from molecular datasets. Molecular eigenspectrums obtained by pre-processing the dataset QM9 were used as the feature maps for prediction. Feature maps for 100 molecules are visualized as examples. The expected outcome of the prediction is a continuous distribution of molecular properties. The prediction can be realized by the traditional analytical method or Machine Learning method. \textbf{b}, The ML methods to predicting molecular properties. The upper portion represents the typical quantum chemistry method. It relies on solving the Schr\"{o}dinger equation approximately for large molecules. This method usually is time consuming. The lower portion represents the proposed machine learning method on a photonic chip. By choosing suitable inputs from a representation of the molecule, a neural network predicts molecular properties by finding complex relationships between the chosen inputs. This is typically faster than the traditional methods.\textcolor{blue}{A typical convolutional neural network model on a digital computer has achieved high prediction accuracy~\cite{pinheiro2020machine}. Here, we focus only on a fully-connected network structure whose training and prediction process are outsourced to photonic chips, as a proof-of-principle of the potential of photonic chips in accuracy and efficiency. The convolutional operation can be achieved on-chip by optical frequency combs~\cite{xu202111,feldmann2021parallel} and microring resonators~\cite{ohno2022si}.}}
    \label{fig:twoways}
\end{figure*}

In this work, we apply an optical neural network chip to the task of implementing a complex-valued neural network to predict multiple quantum mechanical properties of molecules. A diagram of how this process is different from traditional quantum chemistry methods is shown in Figure~\ref{fig:twoways}. Specifically, the contribution of our work is threefold. Firstly, it constitutes the first application of optical machine learning methods to molecular sciences. Secondly, previous works of optical neural networks only consider single-task classification learning, here we extend the optical neural network implementation to a different class of machine learning tasks, namely multi-task regression learning, and obtain satisfactory accuracy. Lastly, we demonstrate that complex-value neural networks, as compared to conventional real-value neural networks, offer superior performance for the prediction of molecular properties.

\begin{figure*}[t]
    \centering
    \includegraphics[width=\textwidth]{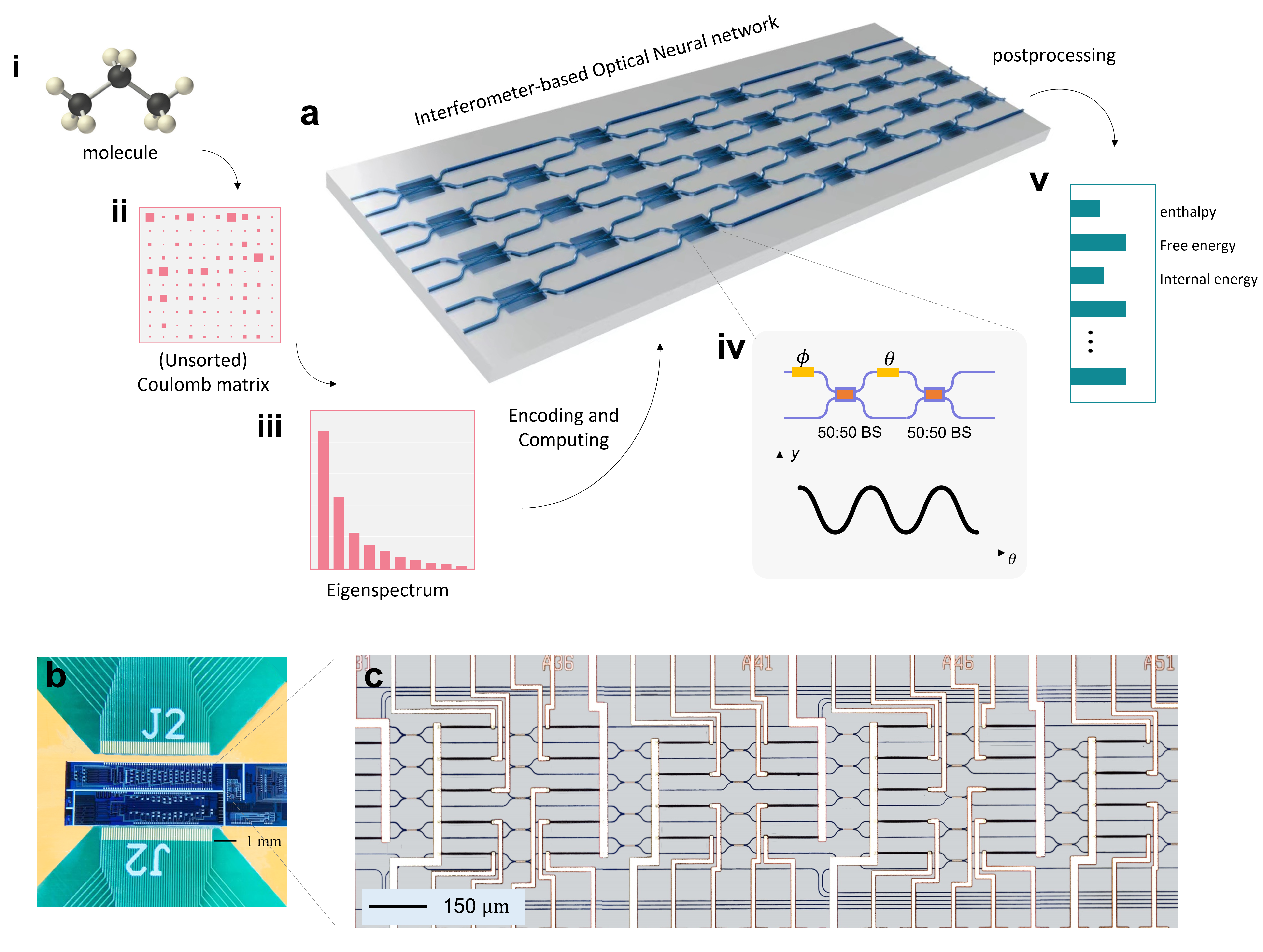}
    \caption{Design and fabrication of the optical chip. \textbf{a}, Architecture of an 8-mode optical neural network chip and the prediction of molecular properties. (i) The representation of a molecule. (ii) Unsorted Coulomb matrix of the molecule. (iii) Eigenvalue spectra representation computed from the Coulomb matrix. (iv) The 2-by-2 optical gate - the fundamental component of the optical neural network chip, and the transmission curve for tuning the internal phase shifter $\theta$. (v) The output of the optical neural network for prediction of molecular properties, including enthalpy, free energy, and internal energy in this demonstration. The optical neural network is divided into a few layers. The first layer is the input layer, which comprises a laser followed by a number of MZIs that distribute the reference light among the waveguides and encode the input into the network by manipulating the phase and magnitude of the light in each waveguide. The next layer comprising the bulk of the ONC is the layer that implements the trainable transformation matrix. The last layer is the output layer which in our cases comprises photon detectors, which enable us to implement the non-linear absolute activation function by detecting the intensity of light in each output waveguide.  \textbf{b}, The packaging of the optical chip. \textbf{c}, A false-color micrograph of the optical chips MZI network with integrated heaters. All 3 layers required for the ONC to function are fabricated on one chip. This chip comprises 8 waveguides and 56 PS for the trainable transformation matrix, allowing us to train up to 8 complex inputs or 16 real inputs. The chip is wire-bonded to a circuit board that provides independent control of each PS by an electronic current driver.}
    \label{fig:chip_images}
\end{figure*}

\section{Molecular Database and Machine Learning Model}

In this work, we use a database comprising computed molecular properties, the QM9 database~\cite{ramakrishnan2014quantum}, to train a photonic neural network. The database contains approximately 134k molecules that consist of up to 9 non-hydrogen heavy atoms (carbon, oxygen, nitrogen, and fluorine). Some notable entries in this database are small amino acids, such as GLY and ALA, and pharmaceutically relevant organic materials, such as pyruvic acid, piperazine, and hydroxyurea. The molecular geometries and chemical properties in the database are calculated by density functional theory methods.

For this dataset, we encoded the chemical structure of the molecule in a Coulomb matrix (see Supplemental Note 1 for details). Since the largest molecules in the database contain up to 29 atoms (including hydrogen), the Coulomb matrix for each molecule is of size $29\times29$, and zero padding is applied to the molecules with fewer than 29 atoms. As we will elaborate on later, our current optical chip support 16 real feature inputs, which are then encoded in 8 complex inputs on the chip. To reduce the dimensionality of the Coulomb matrix to 16, the eigenspectrum representation of the Coulomb matrix is adopted as the feature set. For each Coulomb matrix, we diagonalized it to obtain its 29 eigenvalues and took the largest $16$ eigenvalues as our feature set for each molecule. Our simulations show that even with a restricted subset of the eigenvalues, the dataset is still sufficiently rich to support accurate learning of the chosen chemical properties (\textcolor{blue}{see Figure~\ref{fig:diff_eigenvalues} in} the Supplemental Note). We focus on the predictions of three molecular properties: enthalpy, internal energy, and free energy. More details about our model, such as label scaling and model validation, can be found in Supplemental Notes 2 and 3.

\section{Details of Optical Chip}

Diagrams of the optical chip are shown in Figure \ref{fig:chip_images}a. Similar to a typical neural network implemented on a classical computer, it comprises an input layer, multiple hidden layers, and an output layer. In an optical neural chip (ONC), light signals are encoded and manipulated by controlling their amplitude and phase. The Maxwell equations that describe such systems of interacting waveguides naturally give rise to complex arithmetic. Usually an arbitrarily complex transformation matrix $W$ can be decomposed into $W=V^\dagger DU$ via Singular value decomposition, where $U$ and $V$ are unitary and $D$ is a diagonal matrix. For this work, we consider the simplified transformations that can be expressed as $W=DU$. This is not universal but has a large searching space compared to just a single unitary matrix, and from empirical studies, meets our requirements for the prediction of molecular properties. The processing of input molecular data is shown in Figure \ref{fig:chip_images}a(i)-(v), in which the Coulomb matrix of a molecule is calculated, and then the eigenspectrum of the Coulomb matrix is calculated. Its 16 largest eigenvalues are used and encoded as optical inputs to the chip. The output of the chip is used to predict the molecular properties.

The ONC implements the complex-valued transformation by acting as a multiport interferometer.
Multiport interferometers implement linear transformation between several optical
channels by arranging many Mach-Zehnder interferometers (MZIs) in a specific pattern (a rectangular or triangular mesh)~\cite{clements2016optimal,carolan2015universal,reck1994experimental}, and it can be shown that any arbitrary unitary operator can be decomposed and implemented in such a way. By setting the transmissivity of both BS fixed at $50:50$, each MZI implements the complex unitary transformation between the 2 neighboring mode waveguides indexed by $p$ and $q$ and which can be parameterized by two independent phases $\theta$ and $\phi$:
\begin{gather}
    T_{p,q}(\theta,\phi) =   \left[ {\begin{array}{cccccc}
   1&0&\dots&\dots&\dots&\dots\\
   0&1&\dots&\dots&\dots&\dots\\
   \dots&\dots&e^{i \phi} \sin \theta & e^{i \phi} \cos \theta &\dots&\dots\\
   \dots&\dots& \cos \theta & -\sin \theta&\dots&\dots \\
   \dots&\dots&\dots&\dots&1&0\\
   \dots&\dots&\dots&\dots&0&1\\
  \end{array} } \right],
\end{gather}
The PS can be thermally modulated to tune their phase shifts, allowing us to reconfigure the chip parameters without requiring complicated modifications. In our case, all the PSs are thermally tuned with integrated titanium nitride (TiN) heaters bonded to a PCB. 

\begin{figure*}[t]
    \centering
    \includegraphics[width=0.88\textwidth]{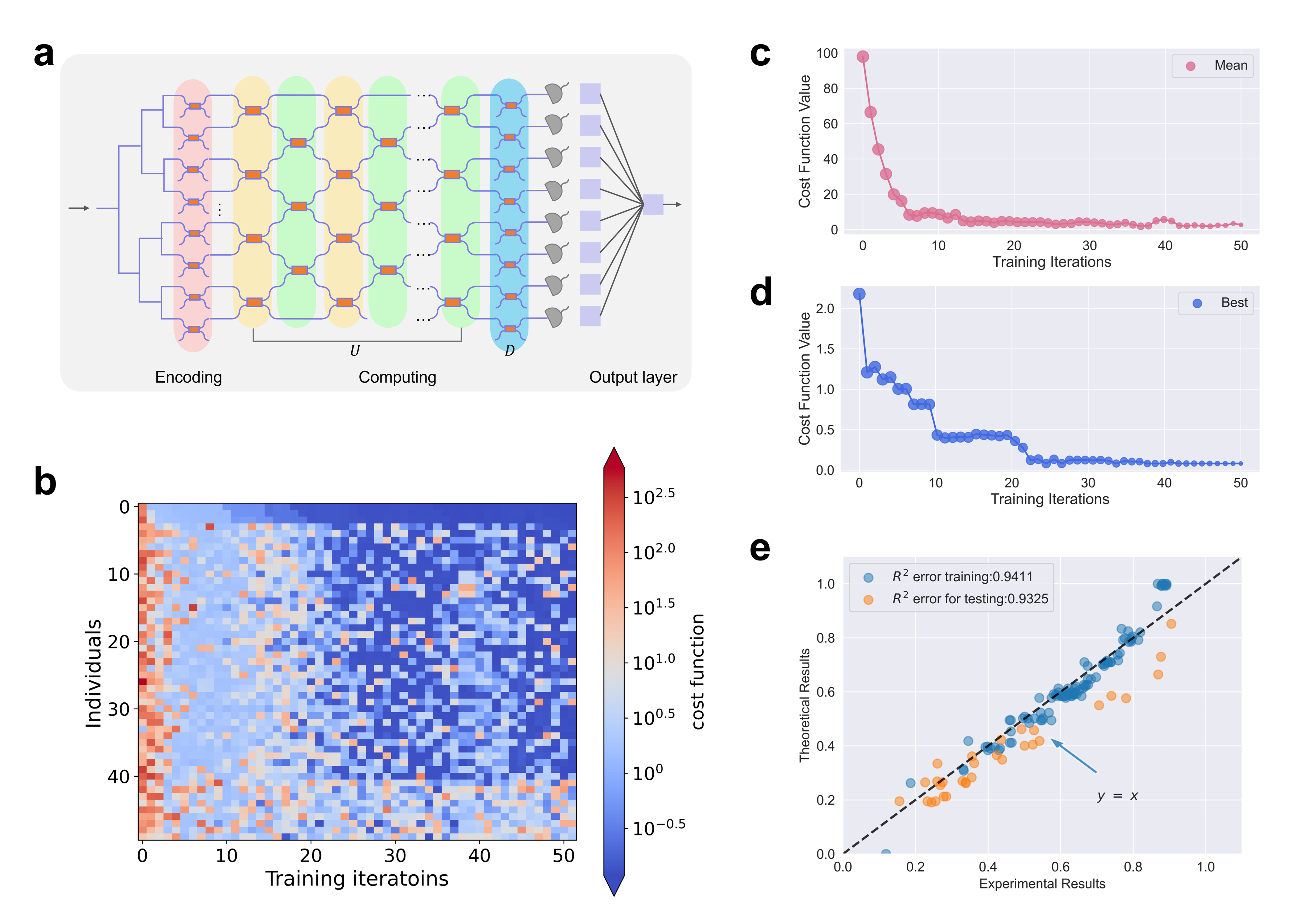}
    \caption{Experimental regression results of the free energy. A total of 100 training instances and 30 testing instances are used for the chip experiment. An initial population of 50 individuals is randomly generated. \textbf{a}, The diagram of the single task learning. Each optical gate is a simplified representation of the MZI structure. \textbf{b}, The visualization of the on-chip training process. The x-axis represents training iterations, \emph{i.e.}, the generations in the Genetic Algorithm. The y-axis represents the 50 individuals in each generation. On the y-axis, rows 1-3 are the elites directly inherited from the previous generation, rows 4-41 are the children created by crossover, and rows 42-50 are children created by the mutation operator. Different colors represent different cost function values, while blue represents smaller cost values and yellow the opposite. From this figure, we can see the evolution of the performance of chip configurations. \textbf{c}, \textbf{d}, The evolution of the mean and the best cost function value in each generation, with increasing generations. \textbf{e}, The regression curve of the trained model on training and testing samples. The 100 training instances (blue data points) are evaluated on the chip after training, to validate whether the training is successful, and the 30 test instances (orange data points) measure the generalizability of the trained model. The regression quality is measured by the coefficient of determination $R^2$ score (the best regression has a $R^2$ score of 1). The training $R^2$ of our trained model on the chip is 0.9411, and the testing score is 0.9325.}
    \label{fig:learning_simulation_results_1}
\end{figure*}

A coherent laser is used to generate the input signals by injecting them into one of the waveguides. The single laser input is split  into 8 equal paths on the chip, each of which can be modulated in both amplitude and phase to encode the input data. After the encoding of the input data, the meshes of the MZIs are next used to implement the optical neural network, or in other words, the trainable transformation matrix. An $N$-mode network transforms the input state into an output state by implementing an arbitrary unitary transformation $U$ with complex weights. $U$ is implemented with multiple rotation matrices $\{T_{p,q}\}$ and a diagonal matrix $D$, $U = \prod_{p=2}^N \prod _{q=1}^{p-1} T_{p,q} D$. $D$ is a diagonal matrix with complex elements with a modulus equal to one on the diagonal, implementing a bias on the outputs, and can be implemented in an interferometer with individual phase shifts on the waveguides at the output end. 
The optical neural chip is thus capable of implementing a complex-valued neural network, which is similar to a conventional real-valued neural network that is implemented on classical computers, with the only difference being that it relies on complex arithmetic. For a system with $N$ waveguides and an arrangement of MZIs described in~\cite{clements2016optimal,carolan2015universal,reck1994experimental}, this has the effect of implementing a hidden layer of $N$ complex-valued neurons. The output of such a neuron
can be expressed as:

\begin{gather}
    y = f\left(\sum_{i=1}^N\omega_i x_i + b\right)\label{eqn:complex_weight},
\end{gather}
where the weights $\omega_i$ and bias $b$ are complex numbers and are determined by the specific choices of $\theta$s and $\phi$s of the MZIs in the ONC.

The light signals at the end of the neural network contain both magnitude and phase information. The detection method (\emph{i.e.}, intensity detection or coherent detection) is related to the activation function that is applicable to the optical output. \textcolor{blue}{In this work, we implement a type of absolute activation function $M(z) = ||z||$. The intensity detection method was chosen because it is experimentally easy to implement, without the need to consume additional MZIs. Coherent detection is possible, but requires two additional columns of MZIs to make light signals from any two adjacent waveguides interfere with each other}, and would lead to the complex Rectified Linear Unit~\cite{brownlee2019gentle} activation function $\text{ModReLU}(z) = \text{ReLU}(||z|| + b)e^{i \theta_z}$. \textcolor{blue}{The detection-based activation function, although not an all-optical implementation, can be extended to multiple layers. For each optical layer, the linear computation is performed optically and the non-linear activation function is realized opto-electronically. For instance, Ashtiani \emph{et al.}~\cite{ashtiani2022chip} demonstrated a photonic deep neural network for sub-nanosecond image classification. Their implementation is first, use a uniformly distributed supply light to provide the same per-neuron optical output range so as to allow scalability to multiple layers without suffering from cascading loss; second, non-linear activation function is realized opto-electronically.}

\textcolor{blue}{In post-processing, a readout layer is applied to linearly combine the light intensities at certain weights to predict molecular properties. Weights in this linear layer are implemented digitally, but are trained concurrently with the trainable parameters of the optical chip layer (see Supplemental Note 10). The training of this linear layer, in addition to extending the network depth and introducing more trainable parameters to improve its capabilities, also fulfills the functions of pre-calibrating the efficiency of photodetectors at different output ports, and scaling the light intensity to the level of molecular property labels.} By increasing the output dimension of this linear map, ONC can perform multi-task learning or the practice of training the network on multiple labels simultaneously.

In our chip, we use an ONC comprising 8 modes and 56 PSs, to simultaneously learn 3 labels (enthalpy, free energy, and internal energy). The input encoding employs additional 15 PSs. This supports up to $16$ real inputs by taking the largest $16$ eigenvalues as our feature set for our data. Our model essentially implements a single complex-valued $8\times 8$ weight matrix representing 8 complex-valued neurons (given in Equation \ref{eqn:complex_weight}), followed by an absolute activation function when reading out.  The Supplementary Information contains results on using a neural network on a digital computer to learn these 3 labels when only given access to the largest $16$ eigenvalues, and they indicate that the $16$ largest eigenvalues are sufficiently rich in information about the underlying system to support accurate learning of these chemical properties. The Supplementary Information also contains comparisons of a similar complex-valued and real-valued neural network on the dataset and suggests that a complex-valued network is superior at learning to a similar real-valued network. 

\begin{figure*}[t]
    \centering
    \includegraphics[width=0.99\textwidth]{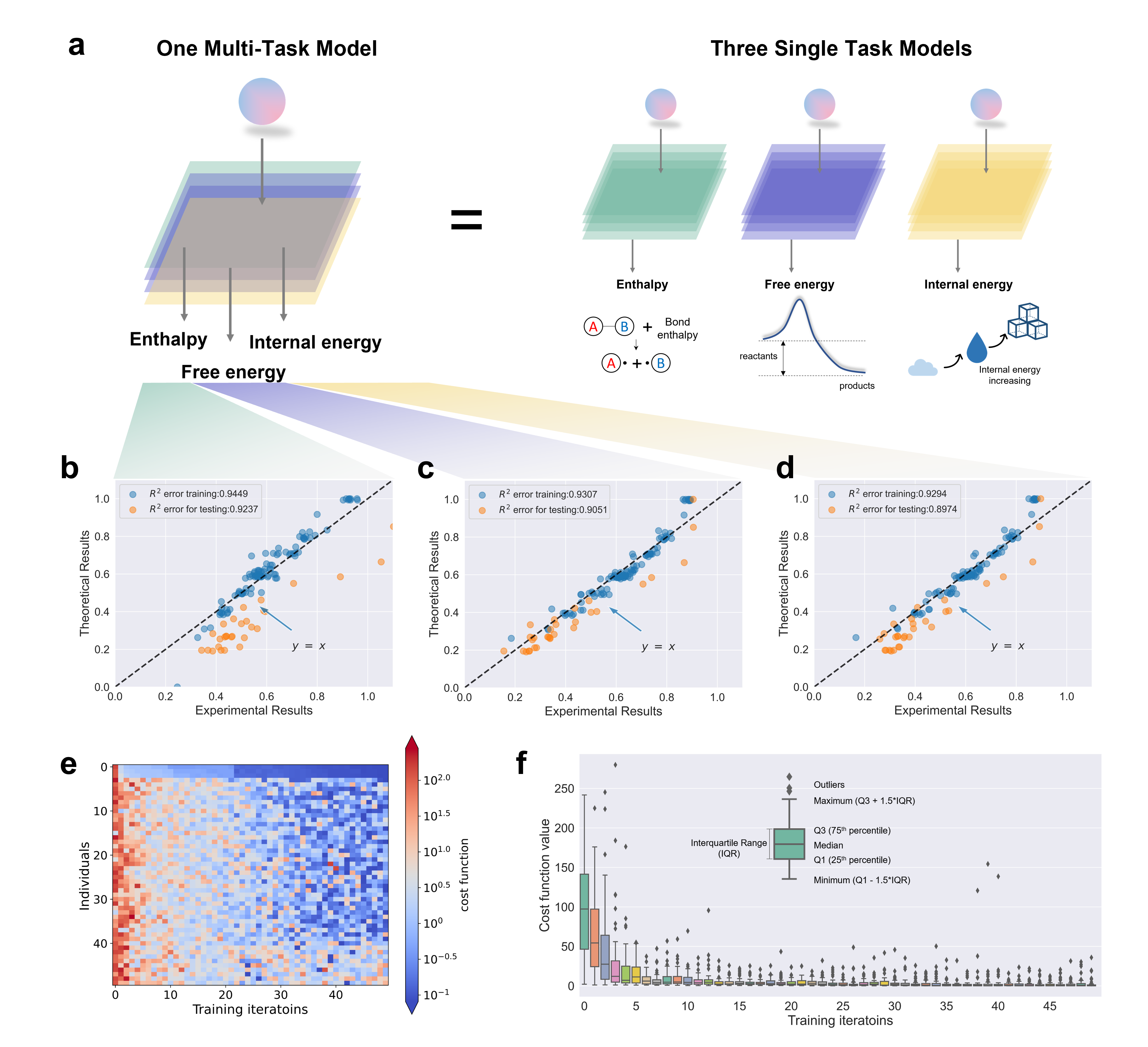}
    \caption{Experimental results were obtained from the chip when regressing on all 3 properties at once (multitask learning). \textbf{a}, A multi-task model, equivalent to three single-task models, is implemented on the chip at one time. Three molecular properties, enthalpy-the amount of energy stored in a bond between atoms in a molecule, free energy-the energy available in a system to do useful work, and internal energy-the total of the kinetic energy due to the motion of molecules and the potential energy associated with the vibrational motion and electric energy of atoms within molecules, are investigated. \textbf{b}, \textbf{c}, \textbf{d}, The regression curves for the properties enthalpy, free energy, and internal energy. \textbf{e}, The visualization of the training process. \textbf{f}, The five-number summary of parameter sets produced in each training iteration. The five-number summary is the minimum, first quartile (Q1), median, third quartile (Q3), and maximum. A convergence trend of the cost function values, as well as the convergence of the variations of all individuals in each iteration, are observed with the training iterations increasing.}
    \label{fig:learning_simulation_results_4}
\end{figure*}

\section{Experimental result}

\begin{table}[t]
\centering
\small
{%
\begin{tabular}{c|ccc}
\hline
 & Enthalpy & Free Energy & Internal Energy \\ \hline
Train MAE & 23.22 & 25.49  & 23.23  \\ \hline
Train $R^2$ & 0.9472  & 0.9411  & 0.9299  \\ \hline
Test MAE & 127.9  & 50.13  & 56.0 \\ \hline
Test $R^2$ & 0.9225 & 0.9325 & 0.9303 \\ \hline
\end{tabular}%
}
\caption{Results from the experimental chip. Labels were learned independently. As can be seen, even for small testing and training data sets, the ONC is able to obtain high coefficients of determination. }
\label{table:separate-training}
\end{table}

First, we discuss the neural networks trained independently for enthalpy, free energy, and internal energy. The number of such macroscopic properties can be increased for a large enough chip. The neural network training process with the genetic algorithm (GA) is illustrated in Fig.~\ref{fig:learning_simulation_results_1}, using free energy as an example. The entire training process is visualized in Fig.~\ref{fig:learning_simulation_results_1}a and shows that the optical neural network is capable of generating good successive generations by crossover (see Supplemental Note 6 for more details). The evolution of the best and mean cost function values with the GA iteration are shown in Fig.~\ref{fig:learning_simulation_results_1}b and c, respectively. It is clear that both values reach plateaus after 20 iterations, indicating the high efficiency of the GA algorithm in optimizing optical neural networks. From the parity plot in Fig.~\ref{fig:learning_simulation_results_1}d, good correlations are seen between the reference and predicted values of free energy for both training and test datasets. 

The quantitative performances (MAE: mean absolute error, and $R^2$: coefficients of determination) of the neural networks for the three thermodynamic properties are summarized in Table \ref{table:separate-training}. Our results demonstrate that the ONC possesses a strong ability to implement a complex-valued neural network model. Already they are able to get impressive coefficients of determination values, and indicate that the model has strong generalization as the coefficients of determinations between the test and train sets are comparable. The prediction on enthalpy has a relatively large MAE, due to the small and thus uneven triaining/testing sets (100/30 samples) randomly selected from the large original dataset volume in this proof-of-princple expeirment. This imperfection in MAE can be imrpoved by having a larger data set.

Table \ref{table:experimental_results_multi} and Figure \ref{fig:learning_simulation_results_4} show the results when we learned the labels concurrently (multitask learning). In the table, we compare the MAEs to a similarly complex neural network on a digital  classical computer, with the same dataset size. Comparable performances are achieved in both training and testing metrics. We also note that our results for the multitask learning are similar to the case where we learned the labels separately. Furthermore, we emphasize that no additional resources were required to perform multitask learning, as the chip inherently implements parallel learning of labels. Although the ONN used in this work and the training and testing data sets are not large-scale, it is a complete proof of principle for the application of photonic chip-based machine learning methods to molecular research, including the training process. The system is the potential to spur on further developments in using ONNs for machine learning tasks.

\begin{table}[htb]
\centering
\small
{%
\begin{tabular}{c|ccc}
\hline
 & Enthalpy & Free Energy & Internal Energy \\ \hline
ONC Train MAE & 27.52 & 25.33  & 23.33  \\ \hline
ONC Train $R^2$ & 0.9449  & 0.9307  & 0.9294  \\ \hline
ONC Test MAE  & 130.70  & 45.67  & 59.06 \\ \hline
ONC Test $R^2$ & 0.9237 & 0.9051 & 0.8974 \\ \hline
DCC Train MAE & 17.34 & 11.90  & 13.38  \\ \hline
DCC Train $R^2$ & 0.9624  & 0.9838  & 0.9812  \\ \hline
DCC Test MAE  & 116.23  & 20.33  & 23.46 \\ \hline 
DCC Test $R^2$ & 0.9399 & 0.9345 & 0.9214 \\ \hline
\end{tabular}%
}
\caption{Results from the optical neural chip (ONC), compared to results from a digital classical computer (DCC). The same data set was used for comparison between the ONC and the digital classical computer. Labels were learned concurrently (multitask learning). The performance is comparable to the digital classical computer on the training set, as well as in generalizing the results for the test set. Both the DCC and ONC were trained with a genetic algorithm.}
\label{table:experimental_results_multi}
\end{table}

\section{Discussions and Conclusions}

The use of machine learning models, in particular deep neural networks, has been demonstrated to be effective in learning the hidden relationships between complex, high-dimensional data sets~\cite{goodfellow2016deep,lecun2015deep}. These techniques provide us with a more efficient and economical way to predict molecular properties and hold enormous promise to accelerate material design and drug discovery.~\cite{xue2016accelerated}. Despite their excellent performance, the complicated structures necessary for such networks, especially for performing more complex learning tasks will require increasingly more computing power and higher energy consumption. 

Furthermore, the vast majority of such models rely on a real-valued neural network. Recent studies suggest that using complex-valued neural networks could significantly improve the performance of similar models~\cite{trabelsi2017deep} by offering richer representational capacity~\cite{reichert2013neuronal}, faster convergence~\cite{arjovsky2016unitary}, strong generalization~\cite{hirose2012generalization} and noise-robust memory mechanisms~\cite{danihelka2016associative}. It has also been shown that complex-valued neural networks have potential in domains where the data is naturally represented with complex numbers, or the problem is complex by design~\cite{bassey2021survey}. Thus, using such complex networks could also potentially be better in dealing with quantum-mechanical problems, as quantum mechanics is inherently a complex-valued theory. However, it is typically not efficient to implement a complex-valued network on a classical digital computer as complex numbers have to be represented by two real numbers on the digital computer~\cite{aizenberg2011complex,yadav2005representation}, which increases the computationally expensive components of the neural network algorithms~\cite{peng2018neuromorphic,sze2017efficient}. 

To overcome this, optical computing has been proposed as an alternate computing platform. The applications of optical computing to run neural networks provide various advantages over classical digital computers, ranging from low electrical power usage, being more energy efficient and robust to noise, and its inherent parallelism allowing us to break down the computation into small steps that are performed in parallel. Most interestingly, it is capable of truly complex-valued arithmetic, allowing us to implement complex-valued neural networks with no additional cost. Such complex-valued networks have been shown to also hold advantages over similar real-valued networks. However, most studies on applying optical computing to neural networks have focused on single-task classification learning. This is a different class of machine learning tasks as compared to what is usually required for applications in chemistry, which is multi-task regression learning, and it was not known previously if such promising prior results in single-task classification would carry over. 

Our work is the first known application of optical neural networks to chemistry. This is the first known application of an optical neural network on chip to optimizing a continuous-valued regression task. It is also the first work to show that such a setup is capable of learning multiple properties at once (multi-task learning) at no additional cost. This also suggests that they would fare well in similar learning paradigms in machine learning, such as transfer learning and multi-output regression. It also demonstrates that complex-valued optical neural networks can achieve performance similar to classical real-valued neural networks implemented on digital computers, for the specific task of learning the chemical properties of molecules. Although currently small, such devices have great prospects for scaling up. It is also reasonable to assume that similar models will be effective in regression tasks for other types of data sets. Such an ONC will be vastly more power efficient than any equivalent neural network run on a DCC, allowing for ever larger neural networks to be trained without a prohibitive energy cost. We give a rough estimate of the power efficiency of our ONC in Supplemental Note 7 and show that already at this scale, it is on orders of magnitude more power efficient than the most power-efficient supercomputers.

By utilizing such optical chips in a manner where we offload the majority of the computational tasks of a neural network off a classical digital computer to them, we also show the potential of creating future hybrid computing systems, blending the advantages of optical and neuromorphic computing with those of classical digital computers.
It is also possible to extend our ONC into a fully-fledged multi-layer neural network with multiple hidden layers by cascading the optical circuits. Such cascaded photonic neural networks have also been demonstrated to be effective in performing various complicated machine learning tasks~\cite{zarei2020integrated}.

\vspace{1em}
\noindent\textbf{\large Acknowledgements}\\
This work is supported by the Singapore Ministry of Education Tier 3 grant (MOE2017-T3-1-001), National Research Foundation grant (MOH-000926), A*STAR research grant (SERC-A18A5b0056), and PUB Singapore's National Water Agency grant (PUB-1804-0082). 

\vspace{1em}
\noindent\textbf{\large Author contributions}\\
H.Z., JWZ.L., L.S. and C.-K.L. jointly conceived the idea. H.Z. and L.W. designed the chip and built the experimental setup. H.C., F.G., X.S.L. and G.Q.L. fabricated the silicon photonic chip. H.Z. and L.W. performed the experiments. All authors contributed to the discussion of experimental results. L.C.K. and A.Q.L. supervised and coordinated all the work. JWZ.L., C.-K.L., H.Z., Y.Z.S, L.K.C., and A.Q.L. wrote the manuscript with contributions from all co-authors.

\vspace{1em}
\noindent\textbf{\large Competing interests}\\
The authors declare that they have no competing interests.


\vspace{1em}
\bibliographystyle{unsrt}
\bibliography{sample}

\clearpage
\newpage
\section*{Supplemental Note}

\renewcommand\thesubsection{\arabic{subsection}}
\subsection{Coulomb matrix}\label{Method:Coulomb}

Much work has been done on figuring out how to efficiently and effectively encode the physical properties of a molecule in a form that can be fed as an input to a machine learning model. One method is to encode it in a form called the Coulomb matrix, and such methods have been routinely applied to the QM7~\cite{blum2009970,rupp2012fast} and QM9~\cite{ruddigkeit2012enumeration,ramakrishnan2014quantum} data sets. While the Coulomb matrix is a low-level molecular descriptor, it is extremely simple to calculate and generate, does not require any domain knowledge, and is already able to obtain promising results~\cite{montavon2013machine,tchagang2019prediction,reddy2021hybrid}.

The elements of the Coulomb matrix are defined by Equations \ref{coulomb1} and \ref{coulomb2}:
\begin{gather}
    c_{ii} = \frac{1}{2} Z_i^{2.4}\label{coulomb1},\\
    c_{ij,i\neq j} = \frac{Z_i Z_j}{|R_i-R_j|}\label{coulomb2},
\end{gather}
where $Z_i$ is the atomic number of atom $i$, and $R_i$ is its position in atomic units. Note that the Coulomb matrix is symmetric and invariant to translation and rotation. The labeling of the atom indexes is also not unique, thus for any given molecule, many different Coulomb matrices could be used to represent it, all related to each other by permuting rows and columns. This is commonly seen as an issue that prevents the Coulomb matrix representation from being used out-of-the-box, as many different Coulomb matrices can be associated with the same molecule. $3$ workarounds are usually used for this~\cite{montavon2012learning}:

\begin{enumerate}
    \item Eigenspectrum representation of the Coulomb matrix. For a $d\times d$ Coulomb matrix, the eigenspectrum representation can be obtained by solving for the eigenvalues of the Coulomb matrix under the constraint $\{\lambda_k\}>0$, where $\{\lambda_k\}$ is the set of eigenvalues. The spectrum $(\lambda_1, \lambda_2, \dots , \lambda_d)$, can then be used as the representation. This also conveniently reduces the dimension of the feature set, although it could potentially remove unrecoverable information about the structure of the molecule.
    \item Sort the Coulomb matrix. We can choose the permutation of atoms whose associated Coulomb matrix $C$ satisfies $||C_i \geq C_{i+1}||$, where $C_i$ denotes the $i^{th}$ row of the Coulomb matrix. By doing so, we ensure that each molecule has a unique Coulomb matrix associated with it.
    \item Extend the data set with randomly sorted Coulomb matrices. This involves generating a set of valid Coulomb matrices for each molecule by randomly permuting rows and columns (equivalent to randomly indexing the molecules) and extending the data set with them. Thus, in the extended data set, each molecule (and thus each set of property labels) has multiple, equally valid, Coulomb matrices associated with it.
\end{enumerate}
All 3 methods have been shown to work quite effectively for regression in neural networks on the QM7 database, which is a smaller subset of the QM9 database which we will be using in this work. 

Furthermore, as mentioned in our main text, our current experimental setup is limited to 16 real feature inputs, encoded into 8 complex inputs. Simulations indicate that the data set still retains enough information about the molecule to permit accurate learning. A comparison of how the accuracy of the machine learning model changes with the number of eigenvalue inputs is shown in Figure \ref{fig:diff_eigenvalues}.

Finally, it is worth mentioning that while the data set provides other properties like the Highest Occupied Molecular Orbital (HOMO) and the heat capacity, from our numerical simulations running similar-sized real-valued and complex-valued neural networks on the data set, we found that if we are limited to the eigenvalue spectrum of the Coulomb matrix for each molecule, the data set was not rich enough to learn those other properties accurately. However, we would like to stress that our optical implementation is general. Any descriptor of the molecule can be used, not just the Coulomb matrix. In our experiment, we utilize the Coulomb matrix as an example, but other, higher dimensional descriptors of the molecule like the molecular matrix and the 3D-MoRSE descriptors could be used too.

\renewcommand*{\thefigure}{S\arabic{figure}}
\setcounter{figure}{0}
\begin{figure*}
    \centering
    \includegraphics[width=\textwidth]{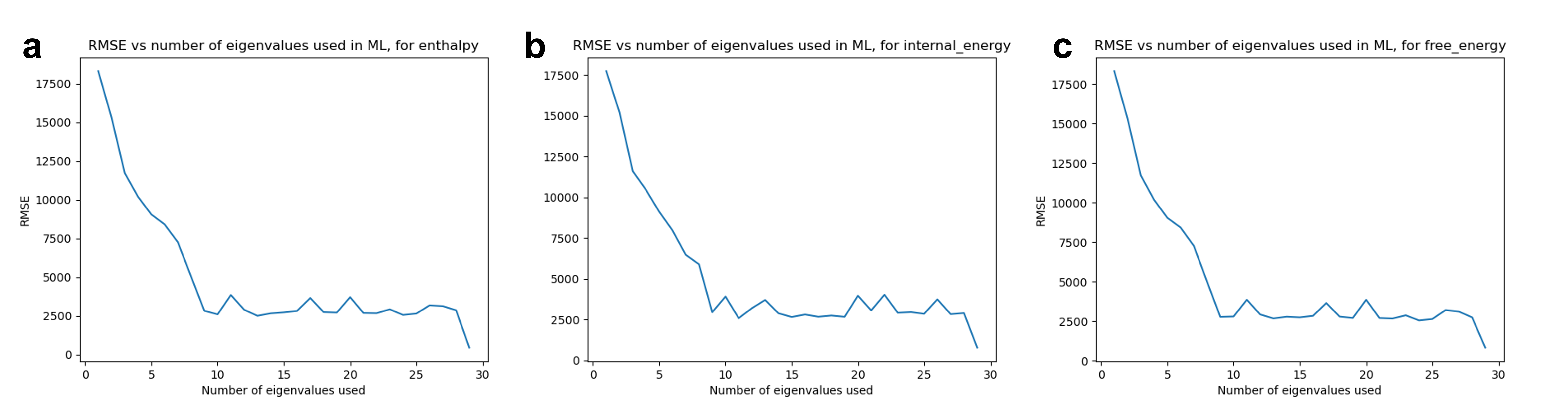}
    \caption{Comparison of a different number of eigenvalues used. Simulations were conducted with a simple model built using PyTorch and assessed using Root Mean Square Error (RMSE). As can be seen, even if restricted to 16 eigenvalues, we are already able to capture most of the details of the system, and using more eigenvalues does not drastically improve the results of the model. This gives us confidence that although our experimental chip limits us to 16 real inputs, we are not giving up too much in accuracy. \textbf{a}, Enthalpy. \textbf{b}, Internal energy.\textbf{c}, Free energy.}
    \label{fig:diff_eigenvalues}
\end{figure*}

\subsection{Scaling of data}

It is known that many machine learning and data-driven analytical methods perform better when the numerical features and labels are scaled to a standard range~\cite{han2011data,ioffe2015batch}. In this work, we are dealing with eigenvalue features that numerically range from a few hundred to single-digit numbers, and property labels that are of the order $10^{5}$. While the input features do not pose too much of a problem due to the way we construct our optical chip, the output labels do need to be scaled to a smaller range. We also want to avoid our labels having negative values, due to the method we are detecting outputs on our optical chip. Thus, we used a simple Min-Max normalization scaling technique on each label separately, so that all the labels are scaled into the range $[0,1]$. This was done with this map:
\begin{gather}
    y_{scaled} = \frac{y-y_{min}}{y_{max}-y_{min}},
\end{gather}
where for each label, $y_{min}(y_{max})$ is the minimum (maximum) value for that label in the entire data set.

\renewcommand*{\thetable}{S\arabic{table}}
\setcounter{table}{0}
\begin{table*}[htb]
\centering
\footnotesize
\begin{tabular}{c|c|c|c|c|c|c}
\hline
                & Enthalpy (R) & Enthalpy (C) & Free Energy (R) & Free Energy (C) & Internal Energy (R) & Internal Energy (C) \\ \hline
MAE (kcal/mol)  & 1197            & 968                & 1208               & 1022                  & 1321                   & 971                       \\ \hline
RMSE (kcal/mol) & 3067            & 2531               & 3065               & 2511                  & 3088                   & 2525                      \\ \hline
Gradient        & 0.9841          & 0.9916             & 0.9848             & 0.9931                & 0.9858                 & 0.9919                    \\ \hline
$R^2$           & 0.9850          & 0.9898             & 0.9850             & 0.9900                & 0.9848                 & 0.9899                    \\ \hline
\end{tabular}
\caption{Results from comparing a real (R) and complex (C) network on the QM9 data. Both networks were constructed with a single hidden layer consisting of 8 neurons, and a ReLU activation function was used in between each layer for both networks. MAE: Mean Absolute Error, RMSE: Root Mean Square Error, Gradient: Gradient of best-fit regression line (= 1 for perfect learning), $R^2$: Coefficient of determination.}
\label{table:real_complex}
\end{table*}

\subsection{Model validation}

For our model, we divided up the data set randomly into a training group consisting of $\frac{4}{5}$ths of the data, and a test group consisting of the remaining $\frac{1}{5}$th of the data. While a more systematic manner of distributing the data could be used (for example, stratified 5-fold cross-validation with identical cross-validation folds, used in~\cite{rupp2012fast}), from our simulations over many random divisions of data, this did not play a major part in affecting the accuracy of our model. We used a simple mean square error objective function for our model and relied on easily available commercial optimization software to train the parameters in our model~\cite{MatlabOTB}.

\textcolor{blue}{
\subsection{The optical gate}
The flow of light is controlled by the 2$\times$2 optical gate, the fundamental element of the programmable optical neural networks. The transformation of the optical gate is unitary. Its implementation as shown in Figure 2(iv) is through a Mach-Zehnder interferometer (MZI) and needs two adjustable $\theta$ and $\phi$ to independently control the power splitting and the relative phase delay. An MZI contains two couplers (which is realized by a multi-mode interference device in our chip) connected by two waveguides, while the optical path can be modulated to introduce phase difference. When the two fixed couplers have a perfect 50:50 split ratio, all coupling ratios from 0\% (bar) to 100\% (cross) are possible. The phase difference is induced by configurable modulators, which are the so-called phase shifters, through the thermo-optic effect. The MZI can be represented as a sequence of BS-PS($\theta$)-BS. Another phase shifter PS($\phi$) is placed at one input port of the MZI to form the optical gate, which is represented by PS($\phi$)-BS-PS($\theta$)-BS. The transformation matrix of BS is
\begin{equation}
\hat{U}_{\rm BS}=\frac{1}{\sqrt{2}}
\begin{bmatrix}
1 &i \\
i &1 \\
\end{bmatrix},
\end{equation}
while the transformation matrix of a phase shifter, for example, the phase shifter $\theta$ on one arm of MZI, can be described as 
\begin{equation}
\hat{U}_{\rm PS}(\theta)=
\begin{bmatrix}
e^{i\theta} &0 \\
0 &1 \\
\end{bmatrix}.
\end{equation}
By sequentially multiplying these separated transformation matrices, the transformation matrix of the optical gate is given by
\begin{equation}
\begin{aligned}
\hat{U}
&=\hat{U}_{\rm BS}\hat{U}_{\rm PS}(\theta)\hat{U}_{\rm BS}\hat{U}_{\rm PS}(\phi)\\
&=\frac{1}{2}
\begin{bmatrix}
1 &i \\
i &1 \\
\end{bmatrix}
\begin{bmatrix}
e^{i\theta} &0 \\
0 &1 \\
\end{bmatrix}
\begin{bmatrix}
1 &i \\
i &1 \\
\end{bmatrix}
\begin{bmatrix}
e^{i\phi} &0 \\
0 &1 \\
\end{bmatrix}\\
&=ie^{i\frac{\theta}{2}}\begin{bmatrix}
e^{i\phi}\sin(\frac{\theta}{2}) &e^{i\phi}\cos(\frac{\theta}{2}) \\
\cos(\frac{\theta}{2}) &-\sin(\frac{\theta}{2})
\end{bmatrix}.
\end{aligned}
\end{equation}
The term $ie^{i\frac{\theta}{2}}$ plays the role of global phase.
}

\subsection{Experimental setup}
The chip is fabricated at Advanced Micro Foundry, Singapore. The chip is wire-bonded to a printed circuit board, providing independent control of each phase shifter by an electronic current driver with 1-kHz frequency and 12-bit resolution (Qontrol Devices, Inc.). Laser pulses are generated by an Ultrafast Optical Clock device (PriTel Inc.) with a repetition rate of 500 MHz, a central wavelength of 1550.12 nm, and a bandwidth of 2 nm. \textcolor{blue}{The input single pump light is coupled into the chip by a one-dimensional subwavelength grating coupler and detected off-chip by 8-mode grating coupler arrays. A polarization controller is utilized to maximize the coupling efficiency of the fiber to chip.} A Peltier controlled by Thorlabs TED200C is used to stabilize the temperature of the chip and reduce the heat fluctuations caused by the ambient temperature and the heat crosstalk within the chip.

\subsection{Genetic algorithm}\label{Method:GA}

For training our neural network, we utilized a genetic algorithm to optimize the parameters for the ONC, to cut down on the time needed to conduct the experiment. 

A genetic algorithm is a global optimization algorithm that can be summarized as so:

\begin{enumerate}
    \item The genetic algorithm first starts out by creating a random initial population. 
    \item The algorithm then iteratively creates a sequence of new populations, known as \textit{generations}. It does so with the following steps:
    \begin{enumerate}
        \item Calculates the scores of the current population with the cost function. These scores are known as \textit{fitness} scores.
        \item Chooses a certain number of the current population with the best \textit{fitness scores}. These are known as the \textit{elite} and are passed on to the next generation.
        \item Produces children from the current generation (parents), with \textit{crossovers} (combining the entries of a pair of parents) and \textit{mutations} (randomly making changes to parents). These children are then also passed on to the next generation.
    \end{enumerate}
    \item The current generation is replaced with the next generation.
    \item This is repeated until the stopping criteria are met. In our case, the stopping criteria is a set number of iterations/generations.
\end{enumerate}

For our genetic algorithm, we conducted it over 50 training iterations (otherwise known as generations), with each generation having a population of 50. Between each generation, we selected the top 3 to survive into the next generation (the elite), then generated 47 additional samples, 39 by crossover, and 8 by mutation. Also due to time constraints, instead of utilizing the whole data set, we only utilized 100 randomly chosen data points in the data set to serve as our training data, and another 30 randomly chosen data points to serve as our test data. This implies that our results will be less accurate than what we expect. However, we believe that even with this small data set, our results demonstrate that our ONC possesses a strong ability to implement a complex-valued neural network model. Useful references can be found in~\cite{holland1962outline,bagley1967behavior,holland1992adaptation,coello2000updated,weise2009global}.

\begin{table*}[t]
\footnotesize
\begin{tabular}{cccccc}
\hline
& \multirow{2}{*}{Network depth} & \multirow{2}{*}{\begin{tabular}[c]{@{}c@{}}No. of parameters \\ (N = 8, M = 3)\end{tabular}} & \multicolumn{2}{c}{Performance} & \multirow{2}{*}{Layer descriptions} \\
 & & & MAE & $R^2$  &  \\ \hline
\begin{tabular}[c]{@{}c@{}}Optical layer \\ + readout layer\end{tabular} 
& 2   & 115    &  17.34   & 0.96   & \begin{tabular}[c]{@{}c@{}}A complex nonlinear layer + a real linear layer\\  $\vec{h}^{N\times 1} = || \vec{W}^{N\times N}\cdot c\vec{x}^{N\times 1}+\vec{b}^{N\times 1}||$\\ $\vec{y}^{M\times 1} =  \vec{R}^{M\times N}\cdot \vec{y}^{N\times 1} + \vec{b'}^{M\times 1}$ \end{tabular}   \\ \hline
Readout layer   & 1  & 27    & 83.20  & 0.31  & \begin{tabular}[c]{@{}c@{}}A real linear layer\\ $\vec{y}^{M\times 1} =  \vec{R}^{M\times N}\cdot \vec{y}^{N\times 1} + \vec{b'}^{M\times 1}$\end{tabular} \\
\hline
\end{tabular}
\caption{Comparison of the network architecture with or without the optical layer. }
\label{tab:wooptical}
\end{table*}

\begin{figure*}[t]
    \centering
    \includegraphics[width=0.66\textwidth]{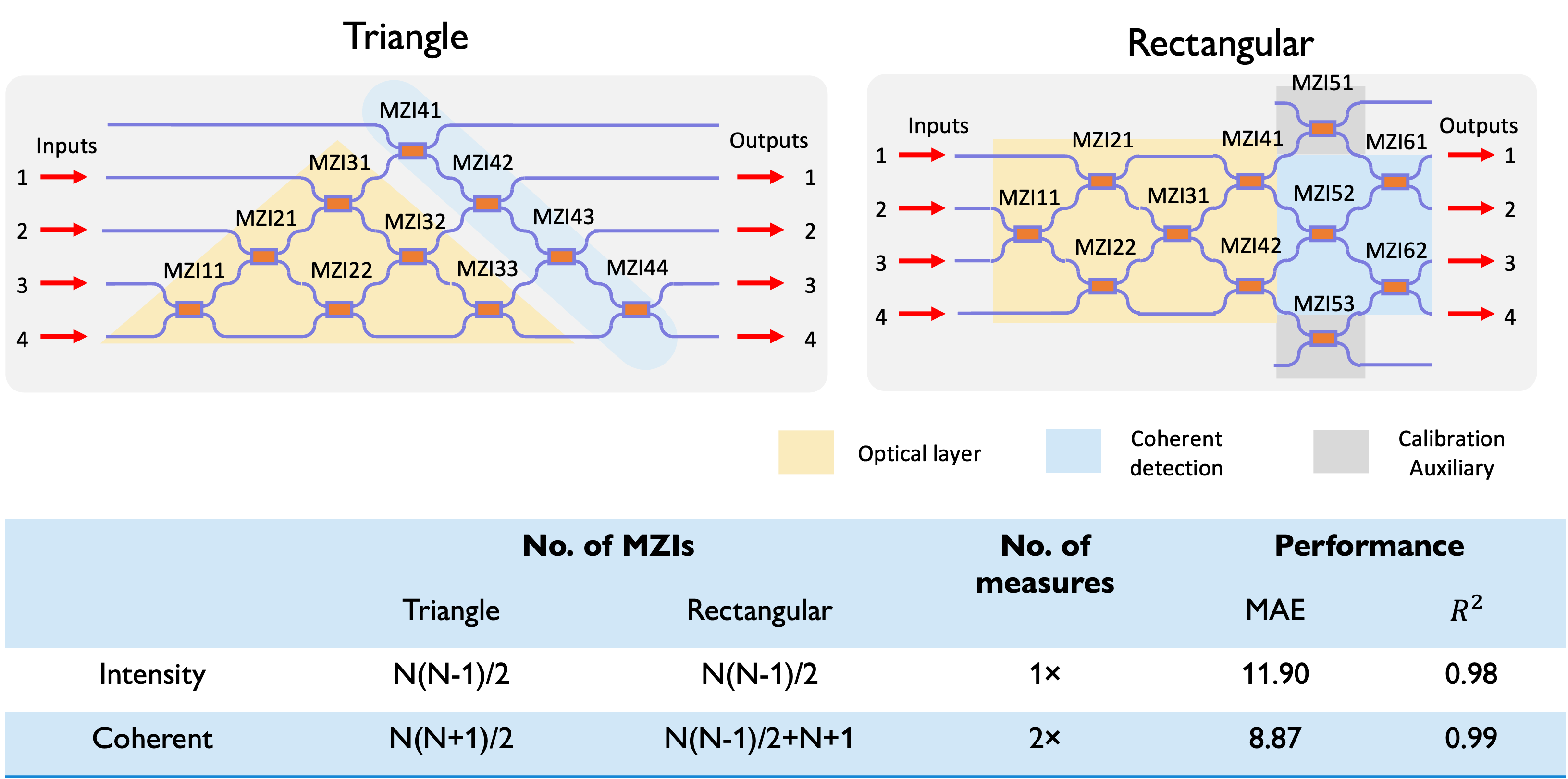}
    \caption{The comparison of coherent detection and intensity detection.}
    \label{fig:detection}
\end{figure*}

\subsection{Resource comparison with digital computers}\label{Method:energy}

In this section we give a rough comparison of how much processing power and power consumption an ONC will use, as compared to training an equivalent network on a digital computer.

For a $N \times N$ matrix, it takes on the order of $O(N^2)$ operations to compute a matrix-matrix product on a digital computer. If we assume that the non-linear activations can be implemented efficiently (typically of order $O(N)$, the number of equivalent floating-point operations per second (FLOPS) on a digital computer an ONC is capable of performing is given by $R$:
\begin{gather}
    R = m*N^2*D \text{ FLOPS},
\end{gather}
where $m$ is the number of layers, and $D$ is the detection rate of the measurement device. The very best photodetectors can potentially achieve detection rates of $\approx 10^{11}$ Hz for single photons~\cite{esmaeil2021superconducting}. For comparison, the very best supercomputers right now have speeds of around $\approx 10^{18}$ FLOPS. It can be seen that it will not take an extremely large optical neural network $N \approx 1000$ to achieve parity with such supercomputers.

If we consider power consumption, the comparison becomes even more favorable for ONCs. If we assume the propagation loss at the photonic circuit level to be negligible (usually on the order of a few percent), the power required to run the ONC is given as $P$:
\begin{gather}
    P = N*p*A \text{ Watt},
\end{gather}
where $A$ is the cross-section of the waveguide, and $p$ is the saturation power (Watt/cm$^2$). For our ONC, $A \approx 10^{-4}$, $p \approx 1$, and $D \approx 10^9$. This gives a power efficiency of $P/R = m*N*10^{13}$ FLOPS/Watt. This is already a few orders of magnitude of power efficiency above the most efficient supercomputers, which are around $50*10^9$ FLOPS/Watt~\cite{top500}. We want to emphasize that the values of $D, p$, and $A$ can all be further improved on, which would only further increase the efficiency.

\begin{figure*}[t]
    \centering
    \includegraphics[width=0.88\textwidth]{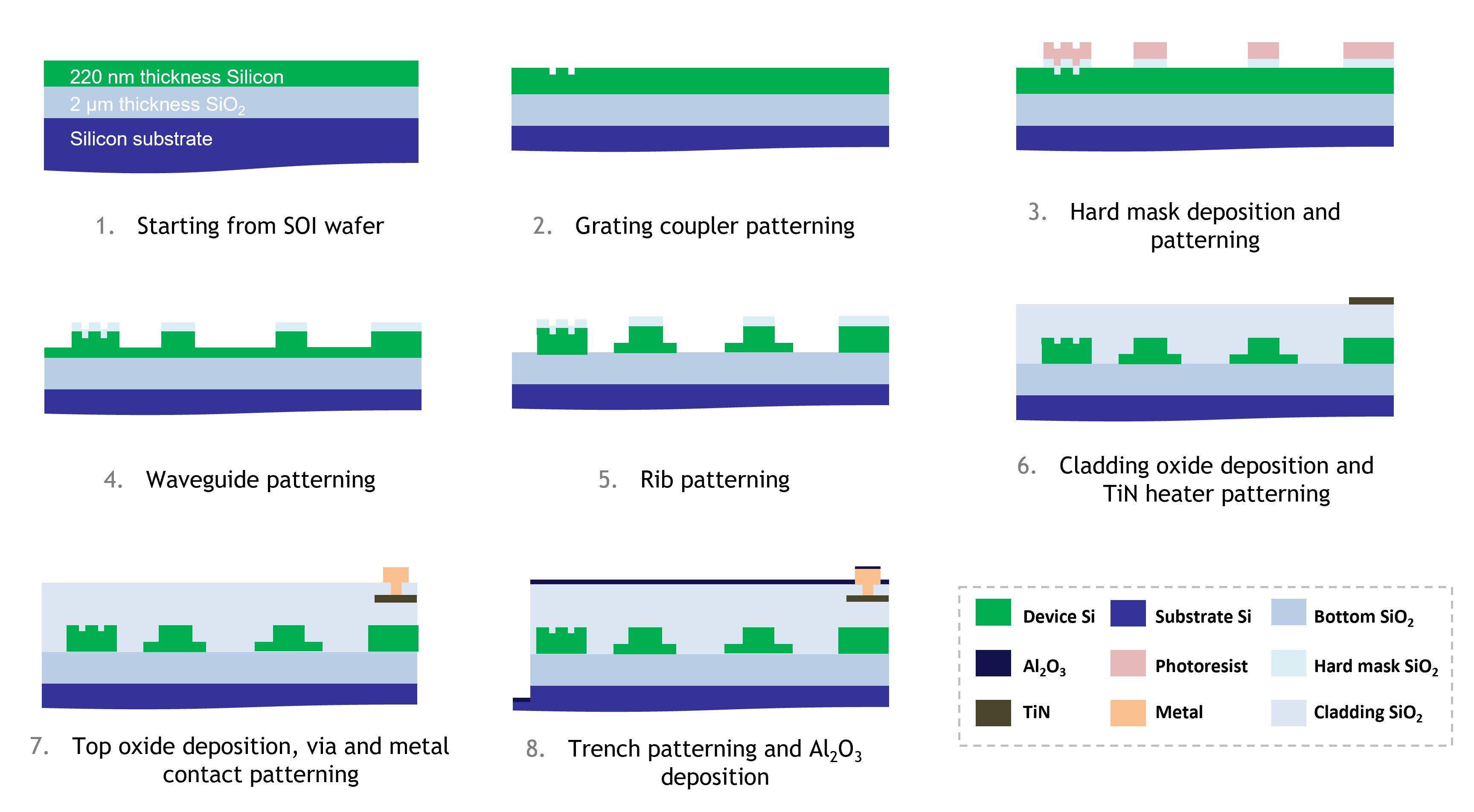}
    \caption{The fabrication process of the integrated silicon photonic chip.}
    \label{fig:fab}
\end{figure*}

\subsection{Comparative studies}\label{si:complexrealcomparison}
\textbf{Comparison of complex-valued and real-valued neural networks.}
Table \ref{table:real_complex} shows the results from simulating similar real and complex networks. As can be seen, in general the complex network performs better, which also implies that to obtain comparable performance, a complex-valued required a smaller chip size and fewer free components (fewer PSs) as compared to a real-valued model on a chip.

\textcolor{blue}{
\textbf{With/without the optical chip layer.}
Table \ref{tab:wooptical} shows the results from simulating the network structure with or without the optical chip layer, from the perspectives of network depth, trainable parameters and performance on free energy (as an example). From the results, the optical chip layer is of great significance to perform the task of predicting molecular properties. In particular, for a single readout layer, it simply performs real-valued linear regression, which is not qualified for tasks like predicting molecular properties.}

\textcolor{blue}{
\textbf{Comparison of coherent and intensity detection.}
Performing coherent detection or intensity detection directly affects the type of nonlinear activation function that can be applied. Figure \ref{fig:detection} shows the results of simulating the same optical neural network with coherent detection or intensity detection. From the results, performing coherent detection has some performance advantages over intensity detection. However, intensity detection is experimentally easier to achieve, using fewer MZIs. For a triangle structure, one additional column is needed by coherent detection; for a rectangular structure, two additional columns are needed. Besides, coherent detection requires measurements of $\sin\theta$ and $\cos\theta$ to fully determine the phase angle $\theta$, thus the number of measurements is doubled.} 

\textcolor{blue}{
\subsection{Fabrication process}
The waveguide structures are built on the common SOI platform, with Si and SiO$_2$ as the core and cladding materials, respectively. The large refractive index contrast between Si ($n_{\rm Si}=3.48$) and SiO$_2$ ($n_{\rm SiO_{2}}=1.44$) ensures the strong confinement of light within the Si core. The waveguide is with cross-sectional dimensions of $500 \times 220$ nm$^2$ to enable single TE mode transmission. Titanium nitride (TiN) is selected as the heater material due to its very high melting point, good electrical conductivity, and CMOS compatibility. The TiN heaters has a length of 150 $\mu$m, width of 3 $\mu$m, and thickness of 120 nm.}

\textcolor{blue}{The fabrication process starts from an 8-inch SOI wafer with 220 nm thick Si device layer and 2 $\mu$m thick buried oxide (BOX) layer. The \textbf{step 1} is to add a laser mark on the wafer for future process tracking. Then the wafer is pre-cleaned with quick dump rinse (QDR) and sulfuric peroxide mix (SPM). After preparing the wafer, the \textbf{step 2} is to pattern the grating coupler structure by deep ultraviolet (DUV) photolithography and reactive ion etching (RIE). The Si device layer is shallowly etched by 110 nm. The \textbf{step 3} is hard mask deposition and patterning. 70 nm thick SiO$_2$ is deposited using plasma enhanced chemical vapor deposition (PECVD) and used as a hard mask for the following Si waveguide etching. The waveguide structure is patterned using another DUV photolithography and RIE steps. In the \textbf{step 4}, the SiO$_2$ hard mask together with the residual photoresist are used for the Si etching to transfer the waveguide pattern from the SiO$_2$ hard mask to the Si device layer. The Si device layer is partially etched by 130 nm. The remaining 90 nm Si is left for the rib structure in step 5. The \textbf{Step 5} is the rib structure patterning. Another lithography step masked the rib region with photoresist. The photoresist together with the hard mask protected the waveguide and rib structure in the final RIE etching step. After this step, all the patterning for the silicon device layer is finished, for different devices such as MZI and grating couplers. In the \textbf{step 6}, SiO$_2$ is deposited, followed by chemical mechanical polishing (CMP) to obtain a planarized 2 $\mu$m thick upper cladding layer. 120 nm thick TiN layer is deposited by physical vapor deposition (PVD). The TiN heater is then patterned. In the \textbf{step 7}, SiO$_2$ is deposited to cover the TiN heater. Another CMP step is performed to realize a 500 nm thick top cladding layer. Then via hole is etched. Finally, a 2 $\mu$m thick Al layer is deposited and the metal contact is patterned. In the last step 8, the wafer is prepared for dicing. The trenches are patterned for sidewall protection and the whole chip surface is covered with Al$_2$O$_3$.}

\textcolor{blue}{
\subsection{The network architecture and hybrid training}\label{Method:Train}
The neural network learns by adjusting its weights and bias iteratively to yield the desired output. The molecular features (network input) and molecular property labels (desired network output) are both encoded on the light intensities. The training of the network finds a mapping/input-output relationship between the network input and output, while the light intensities provide a bridge. In our implementation, the conversion factor from the molecular features (eigenvalues) to the light intensity is a constant, which is denoted as $c$. Suppose the input dimension is $N$, the output dimension is $M$, the network input is $\vec{x}^{N\times 1}$, and the optical chip layer is $\vec{W}^{N\times N}$, then the input light intensity of the optical chip layer is encoded as
\begin{equation*}
    c\vec{x}^{N\times 1}.
\end{equation*}
After applying the optical weight matrix and absolute activation function (by intensity detection) to the input, the output light intensity is
\begin{equation*}
    f(\vec{W}^{N\times N}\cdot c\vec{x}^{N\times 1}+\vec{b}^{N\times 1}),
\end{equation*}
where $f(z)=||z||$, and $\vec{b}^{N\times 1}$ is the bias vector. After the linear output layer $\vec{R}^{M\times N}$, the neural network output is 
\begin{equation}
    \vec{h}^{M\times 1} = \vec{R}^{M\times N}\cdot f(\vec{W}^{N\times N}\cdot c\vec{x}^{N\times 1}+\vec{b}^{N\times 1}).
\end{equation}
The bias of the readout layer is ignored here for simplification. The conversion factors $\vec{D}^{M\times M}$ transforms light intensity $\vec{h}^{N\times 1}$ to molecular properties $\vec{y}^{N\times 1}$ with different physical units, and can be written as a diagonal matrix (since a network output corresponds to a molecular property) as
\begin{equation*}
    \vec{D}^{M\times M}=\begin{bmatrix}
    &d_1 &\cdots &0\\
    &\vdots &\ddots &\vdots\\
    &0 &\cdots &d_M
    \end{bmatrix}.
\end{equation*}
The output intensity after the encoding is
\begin{equation}
\begin{aligned}
\vec{y}^{M\times 1} &= \vec{D}^{M\times M}\cdot\vec{h}^{M\times 1}\\
&= \vec{D}^{M\times M}\cdot  \vec{R}^{M\times N}\cdot f(\vec{W}^{N\times N}\cdot c\vec{x}^{N\times 1}+\vec{b}^{N\times 1})\\
& = (\vec{D}^{M\times M}\cdot  \vec{R}^{M\times N})\cdot f(\vec{W}^{N\times N}\cdot c\vec{x}^{N\times 1}+\vec{b}^{N\times 1}).
\end{aligned}
\end{equation}
In a real implementation, the conversion factors $\vec{D}^{M\times M}$ are combined to the training of the linear output layer as $\vec{R'}^{M\times N} =(\vec{D}^{M\times M}\cdot  \vec{R}^{M\times N})$.}

\begin{figure}[b]
    \centering
    \includegraphics[width=0.5\textwidth]{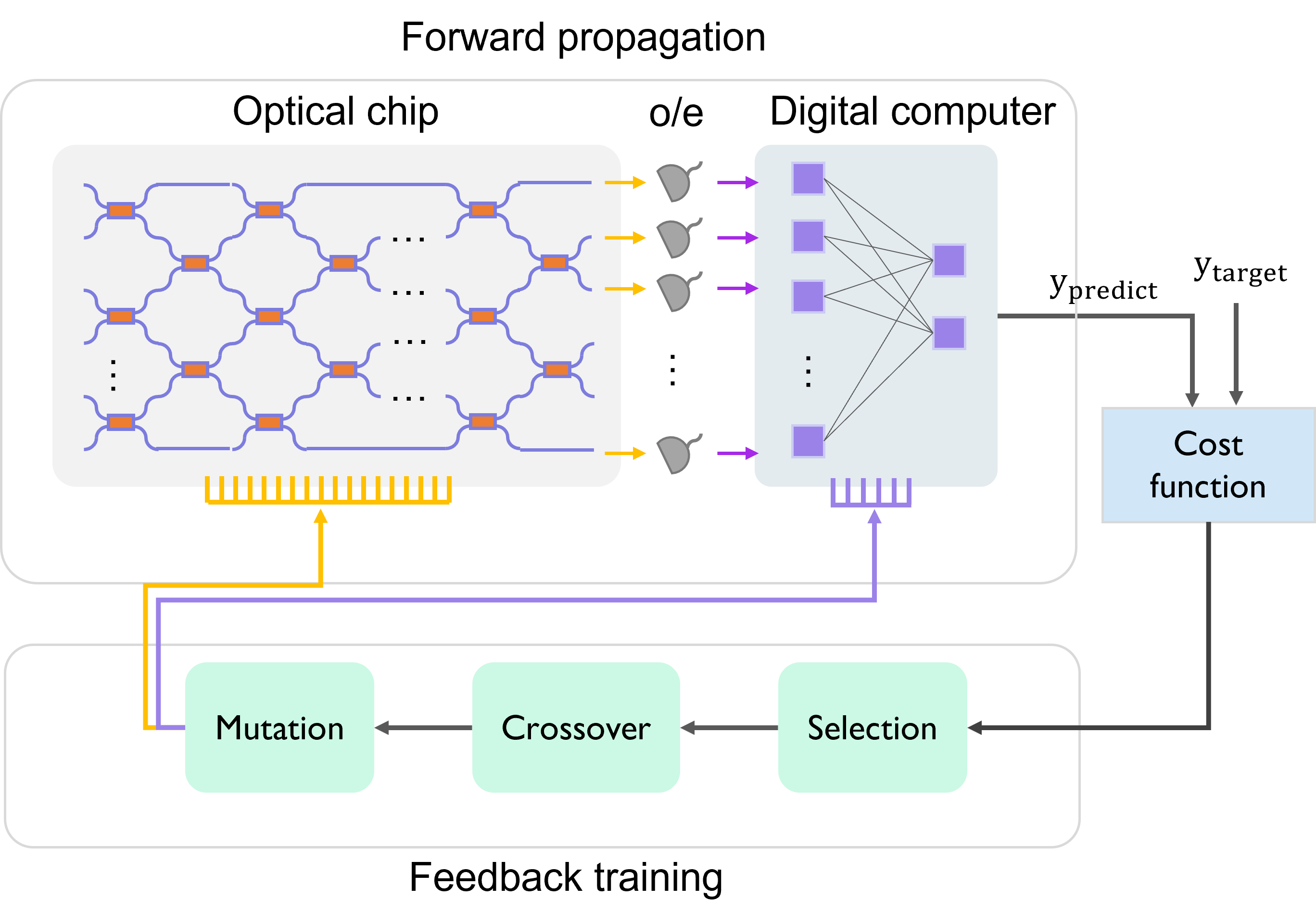}
    \caption{The training flow of the hybrid neural network of an optical chip layer and a digital readout layer. Training parameters in both layer are trained concurrently.}
    \label{fig:trainflow}
\end{figure}

\textcolor{blue}{Overall, the network architecture has two layers, the optical chip layer and the readout layer on a digital computer. The free parameters of the on-chip complex-valued weight matrix $W$, and the free parameters of readout layer $R'$, although implemented on different devices, are trained concurrently. These two layers form a complete forward propagation by feeding the real chip output to the readout layer. Then the final network output is used to compute the cost function, which will be used to decide the directions and step sizes to optimize these free parameters, through an evolutionary optimization algorithm. The training process is depicted in Figure~\ref{fig:trainflow}.}

\end{document}